# Neutron scattering on HoB$_{12}$ - short-range correlations above $T_N$

K. Siemensmeyer,[a] K. Flachbart,[b,*] S. Gabáni,[b] S. Maťaš,[a,b] Y. Paderno,[c]

and N. Shitsevalova,[c]

[a]*Hahn Meiner Institute, Glienicker Str. 100, D-14109 Berlin, Germany*

[b]*Institute of Experimental Physics, Slovak Academy of Sciences, SK-04353 Košice, Slovakia*

[c]*Institute for Problems of Material Science, NASU, UA-252680 Kiev, Ukraine*

**Abstract**

Neutron scattering experiments on frustrated *fcc* - antiferromagnet HoB$_{12}$ have been performed above $T_N$ = 7.4 K. Diffuse scattering patterns indicate that above Néel temperature correlations between neighboring magnetic moments of Ho - ions appear, similar to the appearance of coupled spin-pairs (dimers) in low dimensional magnets. Moreover, the diffuse patterns show a symmetry reduction from *fcc* to *simple cubic*. Analogous behavior in 3D systems is not known, although it was predicted by theory. The role of various interactions leading to this behavior is being analyzed and discussed.

Properties of MB$_{12}$ compounds (dodecaborides), where M denotes a metal ion, have recently attracted considerable interest. This is above all due to a unique combination of their physical properties, such as high melting point, hardness, thermal and chemical stability, and rich electronic and magnetic properties [1 - 3]. They crystallize in the NaCl based *fcc* structure with strong covalent bonds within and between B$_{12}$ clusters. Among them e.g., YB$_{12}$, ZrB$_{12}$ and LuB$_{12}$ are metals which become superconducting at low temperatures, YbB$_{12}$ is a Kondo insulator, and TmB$_{12}$, ErB$_{12}$ and HoB$_{12}$ are metals which order antiferromagnetically in the Kelvin temperature range.

For magnetic dodecaborides it was shown that the indirect exchange interaction of the RKKY type is the dominating mechanism leading to the observed antiferro-magnetic order [4]. Moreover, according to recent specific heat and magnetization measurements on HoB$_{12}$ single crystals and neutron diffraction measurements on powder samples of HoB$_{12}$ [3] it was shown that the magnetic structure of this compound exhibits very complex features. Three magnetic phases were observed below $T_N$ in the *B* vs. *T* phase diagram, and in zero magnetic field the obtained results point to an incommensurate amplitude-modulated magnetic structure [3, 5]. And, it was shown that the dipole-dipole interaction and frustration effects of the fcc lattice, which can be decomposed into corner-sharing tetrahedra, seem to play a important role in the formation of this magnetic structure. In the paramagnetic region, on the other hand, at about 40 K a Schottky contribution to the specific heat of HoB$_{12}$ was observed [1] and interpreted as a manifestation of the crystalline electric field (CEF) on the physical properties above $T_N$. From these results it also follows that the ground state of HoB$_{12}$ is a $\Gamma_5^1$ - triplet. The first and second excited levels are the $\Gamma_3^1$-state at 63 K and the $\Gamma_4^1$-state at 135 K, respectively.

In order to receive more information about the formation of the magnetic structure of HoB$_{12}$, and about the role of various interactions in this process, we investigated the critical scattering of neutrons close to the Néel temperature and the neutron diffuse scattering above $T_N$. In parallel, electrical resistivity and magnetization measurements at low temperatures were performed.

Neutron scattering experiments were performed at the BER II reactor at the Hahn Meitner Institute, Berlin. A 5 T cryomagnet with a variable temperature insert was used to set the temperature and magnet field. Magnetoresistance measurements were performed using a standard *dc* four-terminal

---

method, the magnetization data were taken with a commercial SQUID - magnetometer. Isotopically enriched single crystalline samples of $HoB_{12}$ were prepared by inductive zone melting [3]. All samples were characterized by a Laue picture which has shown *fcc* symmetry from opposite sides of the slabs.

The analysis of neutron scattering patterns close to $T_N$ has shown (Fig. 1), that the critical exponents at these temperatures follow the expected power low dependence. However, the values of the exponents do not match the expectations. Specifically, the line width, which corresponds to the correlation length, behaves as $t^{-\nu}$, with $\nu = 1/3$, where $t = (T/T_N-1)$ is the reduced temperature and $\nu$ is the critical exponent. Landau theory of second order phase transitions expects $\nu = 1/2$, other approximations expect $\nu \geq 1/2$. The same applies for the amplitude of patterns, which can be associated with generalized magnetization, where instead of a $t^\beta$ dependence with $\beta = 1/2$, a dependence with $\beta = 1/3$ was observed.

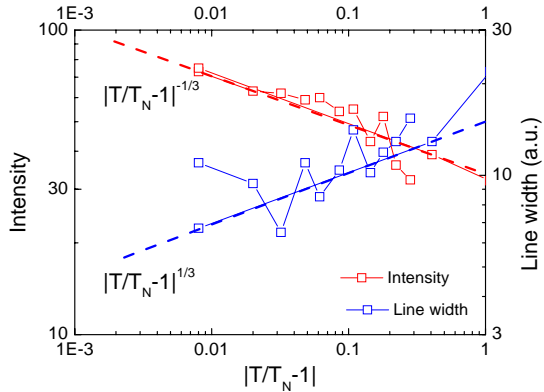

Fig. 1. "Line-width" and "peak-intensity" as a function of reduced temperature.

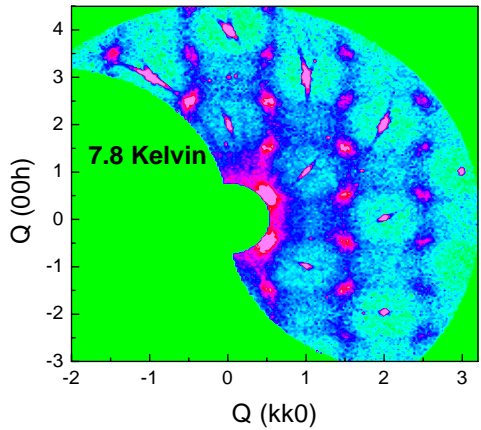

Fig. 2. Diffuse scattering on $HoB_{12}$ above $T_N$.

These results were a stimulus for further investigations of the diffuse scattering (originating from disordered magnetic moments) above the ordering temperature. Fig. 2 shows the diffuse scattering patterns on $HoB_{12}$ at 7.8 K. As follows from these patterns, a strong modulation of diffuse scattering can be seen above $T_N$ (it can be observed also above 77 K). It appears (in Fig. 2 in the form of dark blue parts) at former magnetic reflections, e.g. at (3/2 3/2 3/2), but not at typical crystal structure reflections of the fcc lattice, e.g. at (0 0 2). The modulation of the signal corresponds to a double $MB_{12}$ elementary cell. This feature points to strong correlations between magnetic moments of $Ho^{+3}$ ions placed on face diagonals of the elementary unit, whereas correlations to next neighbors (along (1 1 1) or (1 0 0) directions) are weak. Thus, diffuse patterns show a symmetry reduction from *fcc* symmetry to *simple cubic* symmetry. As this scattering can be resolved also well above $T_N$, it is unlikely to associate it with the critical scattering at Neél temperature. The observed results resemble the appearance of correlated (short range ordered) spin-pairs (dimers) in low dimensional magnets [6]. Analogous behavior in 3D systems is not known, although it was predicted by theory [7, 8]. At $T_N$ the dimers seem to condense into the above aforementioned complex anti-ferromagnetic structure.

It should be noted, that indications for a magnetic anomaly can be seen also in magnetization ($M$) measurements in the paramagnetic state. Here, at about 130 K, a deviation from linearity in $1/M$ vs. $T$ plot can be observed (Fig. 3). This deviation reduces the $M \sim 1/T$ increase of magnetization towards lower temperatures, which may be interpreted as a consequence of short range antiferromagnetic ordering (anti-parallel oriented spin pair formation). As the temperature of deviation correlates with the second excited CEF state, it suggests that the crystalline electric field plays an important role in the formation of this short range ordering. The dipole-dipole interaction, which plays an important role in the ordered state below $T_N$ (its energy is of the order of a few Kelvin), seems to be weak to be considered in the formation of the considered dimer - like structures.

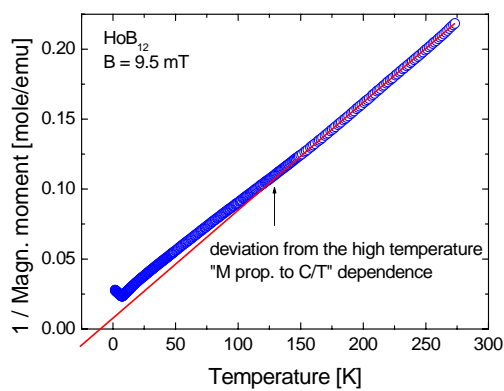

Fig. 3. Inverse magnetization $1/M$ as a function of temperature.

In summary, diffuse neutron scattering patterns indicate that above $T_N$ correlations between neighboring magnetic moments of Ho ions appear. They resemble the appearance of correlated spin-pairs (dimers) in low dimensional magnets. In addition, the diffuse pattern anomalies show a symmetry reduction from *fcc* to *simple cubic*. An analogous behavior in 3D systems is not known, although it was predicted by theory. The dimer - like formation can result from the interplay between CEF and RKKY interactions. However, frustration effects due to the *fcc* symmetry of dodecaborides can also play an important role. To determine exactly the origin of the modulation of diffuse scattering above $T_N$ and the role of particular interactions, further experimental investigations as well as theoretical studies of this class of materials are needed.

The work was supported by the EC through Access to Research Infrastructures of Human Potential Programme under the 6$^{th}$ Framework Program, by the International Association for the Promotion of Cooperation with Scientists from New Independent States INTAS 3036, by the German exchange agency DAAD, by the Slovak Scientific Grant Agencies VEGA, contract 4061, and APVT, contract 0317, and by the contract I/2/2003 of the Slovak Academy of Sciences for the Centres of Excellence.